\documentstyle[12pt]{article}

\begin{document}

\newcommand {\bea}{\begin{eqnarray}}
\newcommand {\eea}{\end{eqnarray}}
\newcommand {\be}{\begin{equation}}
\newcommand {\ee}{\end{equation}}

\def\IR{{\hbox{{\rm I}\kern-.2em\hbox{\rm R}}}}
\def\IH{{\hbox{{\rm I}\kern-.2em\hbox{\rm H}}}}
\def\IC{{\ \hbox{{\rm I}\kern-.6em\hbox{\bf C}}}}
\def\IZ{{\hbox{{\rm Z}\kern-.4em\hbox{\rm Z}}}}

\title{
\begin{flushright}
\begin{small}
hep-th/9805208\\
UPR/807-T \\
May 1998 \\
\end{small}
\end{flushright}
\vspace{1.cm}
The Perturbation Spectrum of Black Holes in $N=8$ Supergravity}
\author{Finn Larsen\\
\small Department of Physics and Astronomy\\
\small University of Pennsylvania\\
\small Philadelphia, PA 19104 \\
\small e-mail: larsen@cvetic.hep.upenn.edu}

\date{ }
\maketitle

\begin{abstract}
The near horizon geometry of four-dimensional black holes 
in the dilute gas regime is $AdS_3\times S^2$, and the global
symmetry group is $SU(2)\times USp(6)$. This is exploited
to calculate their perturbation spectrum using group theoretical 
methods. The result is interpreted in terms of three extreme 
$M5$-branes, orthogonally intersecting over a common string. We also
consider $N=8$ black holes in five dimensions, and compute
the spectrum by explicit decoupling of the equations of motion, extending 
recent work on $N=4$ black holes. This result is interpreted in terms 
of $D1$- and $D5$-branes that are wrapped on a small four-torus.
The spectra are compared with string theory.
\end{abstract}                                
\newpage

\section{Introduction}
\label{sec:intro}

It has been proposed that, in certain limits, the near horizon geometry 
of brane configurations contain all the structure needed to construct 
the underlying quantum theory~\cite{juanads}. In particular, the 
spectrum of classical perturbations in the near horizon region is 
equivalent to that of the quantum operators~\cite{wittenads}, and the 
quantum correlation functions are similarly encoded in the
geometry~\cite{polyakovads,wittenads}. These relations have been
extensively exploited to study conformally invariant gauge theories in four 
and six dimensions. Another interesting system is the bound state
of $D1$-branes and $D5$ branes, because of its relation to black holes in 
five dimensions~\cite{strom96a}, and plausibly to second-quantized
string theory~\cite{dmvv}. In the case of $D1-D5$ the corresponding 
conformal field theory (CFT) is two dimensional, and so the techniques 
for its study are well-developed; moreover, the spectrum of perturbations 
in the near horizon geometry has been computed completely
for $D1-D5$ wrapped on a small $K3$~\cite{sezginads3}. 
The correspondance with the CFT has been strikingly confirmed in some 
instances~\cite{strom98a,martinecads3}; but the map may nevertheless
be imperfect~\cite{vafaads3}.

The purpose of the present paper is to present the spectrum of the 
black holes in $N=8$ supergravity, with emphasis on the
four dimensional case~\cite{dyon,cystrings,hlm}. Their spectrum is 
classified in general, using the global duality symmetries. In the dilute
gas regime the global symmetries are enhanced to $SU(2)\times USp(6)$,
and the near horizon geometry is of the form $AdS_3\times S^2$. 
This structure is exploited to give a derivation of all the conformal
weights and the associated global quantum numbers, using group 
theoretical methods. The result provides the starting point for a 
more detailed study of the conformal field theory that describes the 
internal structure of black holes in four 
dimensions~\cite{dvv2,witt4d,vafa4d}. 
This theory is not well-understood from fundamental string theory 
so the spectrum of perturbations gives new results, such as the complete 
list of chiral primary fields.

The spectrum of black hole perturbations can be applied to the study 
of the dynamics underlying the emission and absorption of Hawking 
radiation. In the simplest case, the minimally coupled scalar field,
there is a quantitative microscopic model that gives both the 
amplitude~\cite{mathur}, and the energy dependence of the Hawking 
radiation~\cite{greybody}, in perfect agreement with semiclassical 
calculations of greybody factors.
For more complicated processes there are interesting results, including
~\cite{cgkt,greybody2,krasnitz97,intscalar,mpartial,gubser,hosomichi,cl97d}, 
indicating that most or all processes can be interpreted as interactions 
in an effective string theory. The spectrum of black hole perturbations, 
presented in this paper, gives the quantum numbers of the black hole 
constituents, whose collisions give rise to the Hawking radiation. 
However, the detailed picture of the black hole dynamics remains incomplete: 
on the semi-classical side, it is not in general clear how the near-horizon 
wave functions match onto the free wave functions in the asymptotic 
Minkowski space. Microscopically, the couplings between operators in the 
conformal field theory and the fields in the bulk of spacetime are not 
yet understood in detail. And, perhaps most seriously, the present
discussion appears to apply only for black holes in the ``dilute gas'' 
regime. Despite these caveats, the results presented here provide 
progress towards a theory of black hole dynamics.

In the above, the discussion has been framed in terms of the properties 
of black holes in four dimensions. However, in the dilute gas regime 
the spacetime is effectively that of a black string in five dimensions. 
This string can be interpreted as the intersection manifold of three 
intersecting $M5$-branes. The interpretation as a regular black hole in 
four dimensions requires that the effective string wraps a compact 
dimension, and momentum must be added that flows along the string, in 
one or both directions. However, in the near horizon region, the black 
hole spacetime and that of three extreme $M5$-branes, uncompactified 
along their common string, differ only in their global properties. Thus, 
since the present paper studies {\it local} properties of spacetime, the 
two perspectives are equally valid. Further discussion of the relation 
between the two interpretations was given, in the context of rotating 
black holes in five dimensions (or rotating black strings in six 
dimensions), in~\cite{cl98a}.

The $AdS/CFT$ correspondance has been exploited extensively 
to study black holes in five 
dimensions~\cite{strom98a,martinecads3,deboer} 
(see also~\cite{btzentropy,teo,myung}). Most of this work has been in 
the context of the black holes in $N=4$ supergravity, interpreted 
microscopically as $D1-D5$ wrapped on a small $K3$ manifold that is 
transverse to the $D1$, and within the $D5$. In sec.~\ref{sec:5d} we 
discuss how these results are extended to black holes in $N=8$ 
supergravity, or $D1-D5$ wrapped on a small transverse four-torus 
$T^4$. The perturbation spectrum is computed by explicit decoupling 
of the linearized wave equations satisfied by the bosonic perturbations, 
as an illustration of those methods.

This paper is organized as follows. In section~\ref{sec:4d} 
we consider black holes in four dimensions. First, in sec.~\ref{sec:global},
we analyze the global symmetries, considering in turn general black holes, 
extreme black holes, and dilute gas black holes.
Then, in sec.~\ref{sec:supmul}, the superconformal multiplets of
perturbations around the dilute gas black holes are calculated, 
using indirect arguments; finally, in sec.~\ref{sec:chiral4d}, the 
complete list of chiral primaries is presented. Section~\ref{sec:5d} 
concerns five-dimensional black holes, extending results previously 
derived for 
$N=4$ black holes to the case of $N=8$ black holes. In this section
the spectrum is computed explicitly, by decoupling of the equations
of motion. In the concluding section~\ref{sec:string}, we compare the 
spectrum of perturbations with that of the underlying conformal field 
theory. The five dimensional case is presented in some detail, and the 
four dimensional black holes are commented on.

\section{$N=8$ Black Holes in Four Dimensions.}
\label{sec:4d}
\subsection{Global Symmetries}
\label{sec:global}
We are interested in the perturbation spectrum of the background 
specified by 3 orthogonally intersecting $M5$-branes. It is assumed that the 
configuration is compactified along all dimensions within the $M5$-branes, 
leaving $4$ noncompact dimensions. In practice, additional dimensions
can be decompactified, as long as the background has been averaged over 
the ``internal'' directions; the global specification of spacetime
will not enter our considerations. The configuration can be interpreted
in terms of black holes in four dimensions and we begin the discussion by 
considering the most general black holes in four dimensions. This
generality will give some additional properties that are not strictly 
needed here, but they are of interest in their own right.

\paragraph{General considerations:}
The duality group of $N=8$ supergravity in four dimensions is
$E_{7(7)}$~\cite{cremmer79}. Any choice of vacuum configuration breaks 
this global symmetry spontaneously to its maximal compact subgroup $SU(8)$,
due to the specification of the scalars at infinity. The supersymmetry
generators $Q_\alpha^A$ transform in the fundamental of $SU(8)$,
and thus the particle spectrum in the four dimensional Minkowski 
vacuum is $70$ scalars $P^{ABCD}$, $56$ Weyl fermions $\psi_\alpha^{ABC}$,
$28$ vectors $F^{AB}_{\mu\nu}$, $8$ gravitini $\psi_{\alpha\mu}^{A}$, 
and the graviton $G_{\mu\nu}$. In each case the $SU(8)$ indices
$A=1,\cdots,8$ are fully antisymmetrized.

The global symmetry is in general broken further in a nontrivial 
background. This is most conveniently analyzed in terms of the central 
charge matrix ${\cal Z}_{AB}$, an antisymmetric tensor that transforms 
in the ${\bf 28}$ of the $SU(8)$. It can be represented up to an $SU(8)$ 
transformation as an antisymmetric $8\times 8$ matrix with the 
skew-eigenvalues (see~\cite{kallosh96a,hull96} and references therein):
\bea
{\cal Z}_{12}&=&{\bf Q_1}+{\bf Q_2}+{\bf Q_3}+{\bf Q_4},\\
{\cal Z}_{34}&=&{\bf Q_1}+{\bf Q_2}-{\bf Q_3}-{\bf Q_4},\\
{\cal Z}_{56}&=&{\bf Q_1}-{\bf Q_2}+{\bf Q_3}-{\bf Q_4},\\
{\cal Z}_{78}&=&{\bf Q_1}-{\bf Q_2}-{\bf Q_3}+{\bf Q_4}~.
\eea
In this formula the boldfaced symbols denote the physical (dressed) 
charges. These are the charges that appear in the spacetime 
solution; they are related to the quantized microscopic charges 
by multiplication with moduli and dimensionful constants. The central 
charge matrix generically breaks the duality group 
$SU(8)\rightarrow SU(2)^4$. This organizes the scalar 
fields $P^{ABCD}$, transforming in the ${\bf 70}$ of $SU(8)$, into 1
$({\bf 2},{\bf 2},{\bf 2},{\bf 2})$, 
2 $({\bf 2},{\bf 2},{\bf 1},{\bf 1})$ ( + $2\times 5$ permutations) 
and 6 $({\bf 1},{\bf 1},{\bf 1},{\bf 1})$. Similar results are easily 
derived for the fields with higher spin.

The symmetry breaking pattern determines the wave equations satisfied 
by the individual fields, {\it via} the coset 
construction (see {\it e.g.}~\cite{freN2}), and 
this gives the explicit connection to much work on greybody factors.
We will not need these details, but let us note that the symmetry 
breaking pattern for the scalars show that there are exactly $16$ 
minimally coupled scalars, $48$ intermediate scalars (in 6 different 
varieties), and $6$ fixed scalars. Furthermore, the spacetime parities
are such that, for each type of scalar, there are equally many proper 
scalars and pseudo-scalars.

\paragraph{Extreme black holes:}
For {\it extreme} black holes the dynamics determines the moduli 
in terms of the microscopic charges so that, 
{\it in the near horizon region}, the physical charges are 
identical~\cite{kallosh96b}:
\be
{\bf Q_1}={\bf Q_2}={\bf Q_3}={\bf Q_4}~.
\label{eq:cequal}
\ee
Then three of the skew-eigenvalues of the central charge matrix vanish 
and the global symmetry is broken $SU(8)\rightarrow SU(2)\times SU(6)$. 
The vacuum multiplets decompose as:
\bea
\begin{array}{ccc}
                     SU(8)& SU(2)\times SU(6)&\\
{\bf 70}  &({\bf 2},{\bf 20})\oplus 2({\bf 1},{\bf 15}) & P^{ABCD}  \\
{\bf 56}  &({\bf 1},{\bf 20})\oplus ({\bf 2},{\bf 15})
                       \oplus ({\bf 1},{\bf 6}) & \psi_\alpha^{ABC}   \\
{\bf 28}   &({\bf 1},{\bf 15})\oplus ({\bf 2},{\bf 6})
                      \oplus ({\bf 1},{\bf 1}) & F^{AB}_{\mu\nu}     \\
{\bf 8}  &({\bf 2},{\bf 1})\oplus ({\bf 1},{\bf 6}) &
\psi_{\alpha\mu}^{A} \\
{\bf 1} & ({\bf 1},{\bf 1}) & G_{\mu\nu}   
\\
\label{eq:extfields}
\end{array}
\eea
In the extreme case the background preserves $N=1$ SUSY which, in the 
horizon region, is enhanced to $N=2$ SUSY. Therefore the fields
eq.~\ref{eq:extfields} can be represented as multiplets of $N=2$ SUSY. 
The supercharges transform according to the 
$({\bf 2},{\bf 1})$ of the $SU(2)\times SU(6)$, so the various components 
of the supermultiplet transform differently under $SU(2)$;
however, a single $SU(6)$ quantum number can be assigned to the
complete $N=2$ multiplet. We can therefore organize
table~\ref{eq:extfields}
as:
\bea
\begin{array}{ccc}
   {\rm multiplet} & {\rm content} & SU(6)\\
{\rm Hyper} & 2S + F & {\bf 20}  \\
{\rm Vector} & 2S + 2F + V & {\bf 15} \\
{\rm Gravitino}& Gi + 2V + F & {\bf 6} \\
{\rm Graviton }& G + 2Gi + V & {\bf 1}\\
\label{eq:N2multi}
\end{array}
\eea
The graviton, the gravitino, the vector, the Weyl fermion, and
the scalar are denoted $G$, $Gi$, $V$, $F$ and $S$, respectively. 
In four dimensions each field has two degrees of freedom, except
the scalar that has only one. With these degeneracies, the numbers 
of bosons and fermions in each multiplet agree.

Since supersymmetry is broken to $N=2$, the gravitini in the
${\bf 6}$ are {\it massive}; in particular, the Weyl fermions in
the gravitino multiplet are in fact absorbed by the gravitini to 
form the physical components with a longitudinal vector index.

In the extreme limit, the intermediate scalars of the general 
classification combine with the minimal and the fixed scalars, forming 
larger multiplets under the enhanced symmetry group. In the near horizon 
region of an extreme black hole the $40$ scalars in the hyper-multiplets 
are all minimally coupled, because $({\bf 2},{\bf 20})$ contains 
$({\bf 2},{\bf 2},{\bf 2},{\bf 2})$ in its decomposition under
$SU(2)^4$. Similarly, the $30$ scalars in the vector-multiplets are all
fixed scalars, because $({\bf 1},{\bf 15})$ 
contains $({\bf 1},{\bf 1},{\bf 1},{\bf 1})$ in its decomposition under
$SU(2)^4$. These result agree with those of~\cite{frefixed}.

\paragraph{Near extreme black holes:}
Next, consider near extreme black holes in the ``dilute gas'' regime.
Since the supersymmetry is not restored in the horizon region 
the ``fixed'' scalars do not take on precisely their fixed point 
values; however some of them do, so that:
\be
{\bf Q_1}\neq {\bf Q_2}={\bf Q_3}={\bf Q_4}~.
\label{eq:dilgas}
\ee
This equation can be verified explicitly in specific examples; more
generally it can be taken as a duality invariant {\it definition}
of the dilute gas regime. Eq.~\ref{eq:dilgas} implies that three of 
the eigenvalues of the central charge matrix are identical, without 
being zero. Therefore the global symmetry is broken as 
$SU(8)\rightarrow SU(2)\times USp(6)$. 

The $SU(6)$ multiplets that decompose nontrivially under $USp(6)$ are:
\bea
{\bf 15}\rightarrow {\bf 14}\oplus {\bf 1}~,
\label{eq:15r} \\
{\bf 20}\rightarrow {\bf 14^\prime}\oplus {\bf 6}~,
\label{eq:20r}
\eea
where the ${\bf 14}$ is the antisymmetric 2-tensor of $USp(6)$, 
and the  ${\bf 14^\prime}$ is the antisymmetric 3-tensor. In both 
cases traces are removed, making it possible that the two representations
have the same dimension without being equivalent. It follows by
comparing eqs.~\ref{eq:15r} and~\ref{eq:20r} with table~\ref{eq:N2multi}
that the breaking from $SU(6)$ to $USp(6)$ only affects the hyper-multiplet 
and the vector-multiplet, dividing each of them into two smaller
multiplets. 
In the dilute gas regime one of the internal dimensions effectively 
decompactifies; so the local Lorentz group is enhanced from $SO(3,1)$ 
to $SO(4,1)$. As result, the $D=4$ graviton  multiplet combines with 
the ``little'' vector-multiplet following from eq.~\ref{eq:15r}, 
in the ${\bf 1}$ of $USp(6)$, and form a single $D=5$ graviton
multiplet. Similarly, the $D=4$ massive gravitino combines with the 
``little'' hyper-multiplet following from eq.~\ref{eq:20r}, in the 
${\bf 6}$ of $USp(6)$, and form a $D=5$ massive gravitino multiplet. 

In summary, the structure of the supermultiplets for perturbations
of black holes in the dilute gas regime, is:
\begin{itemize}
\item
The $N=2$ graviton multiplet has 1 graviton, 2 gravitini (in the
${\bf 2}$ of the global $SU(2$) ), and 1 vector. 
They have $5$, $2\times 4$, 
and $3$ degrees of freedom, respectively. The multiplet is ${\bf 1}$ 
under the global $USp(6)$. 

\item
The massive gravitino multiplet has 1 massive gravitino and
2 vectors (in the ${\bf 2}$ of the global $SU(2)$ ); they have $6$ and
$2\times 3$ degrees of freedom, respectively. The multiplet is ${\bf 6}$ 
under the global $USp(6)$. 

\item
The vector-multiplet has 1 vector, 2 fermions (in the ${\bf 2}$
of the global $SU(2$) ), and 1 scalar. They have $3$, $2\times 2$, 
and $1$ degrees of freedom, respectively. The multiplet is ${\bf 14}$ 
under the global $USp(6)$. The scalars are the fixed scalars.

\item
The hyper-multiplet has 1 fermion and 2 scalars (in the ${\bf 2}$
of the $SU(2)$). They have $1\times 2$ and $2\times 1$, degrees of 
freedom, respectively. The multiplet is ${\bf 14^\prime}$ under the global 
$USp(6)$. The scalars are the minimally coupled scalars.

\end{itemize}

The classification according to global symmetries is complete as
it stands, but the background in general breaks Lorentz invariance.
Thus the wave function of the various components of the $N=2$, $D=5$
multiplets are not in general related in any simple way.

Duality invariance is manifest in the discussion\footnote{In 
particular, the classification persists for the most
general black holes, with five independent charges~\cite{cfthair1}.
In fact, the near horizon geometry of these black holes is identical to
the one considered here, when the dilute gas condition 
(eq.~\ref{eq:dilgas}) is satisfied.}; no specific higher dimensional 
interpretation of the configuration has been assumed. However, some 
configurations behave simpler under five dimensional boost invariance, 
leaving the organization into $D=5$ multiplets more useful. In the 
remainder of this work we consider the three orthogonally intersecting 
$M5$-branes.

The multiplet structure given above can be recovered directly from the
five-dimensional perspective, by considering the near-horizon
geometry of an extreme black hole; the computation is similar 
to the extreme case in four dimensions, except that the compact 
duality group in five dimensions is $USp(8)$.

The extreme black holes and the dilute gas black holes are characterized
by eq.~\ref{eq:cequal} and eq.~\ref{eq:dilgas}, respectively.
The boundary conditions on the scalars at infinity can be chosen so
that these equations are maintained throughout spacetime\footnote{In the
extreme case such black holes are referred to as double extreme 
black holes (see {\it e.g.}~\cite{kallosh96c}).}. With this choice of 
moduli the global symmetries, respectively $SU(6)$ and $USp(6)$, can be 
applied away from the near-horizon region. Thus these symmetries
classify the full greybody factors, rather than just the near-horizon 
wave functions. 

In the case of three orthogonally intersecting $M5$-branes, it is 
possible to include momentum running along the line of intersection (in 
both directions, breaking supersymmetry). The geometry of this
configuration is ${\rm BTZ}\times S^2\times T^6$ with the effective BTZ 
mass and angular momentum dependent on the momentum and the energy
above extremality, as given in~\cite{bl98}. However, the BTZ black
hole is locally $AdS_3$ and so the local geometry remains 
$AdS_3\times S^2\times T^6$ after this apparent generalization, leaving 
the spectrum of perturbations unaffected. Thus, for local properties,
the possibility of momentum can be ignored without loss of generality. 
In particular, the spectrum is organized in supermultiplets, even when 
supersymmetry is broken.

\subsection{Superconformal multiplets}
\label{sec:supmul}
In this section we find the perturbation spectrum of 
the near horizon region of three intersecting $M5$ branes, as classified 
under the $(4,0)$ superconformal symmetry of the underlying CFT in two 
dimensions. This is accomplished by an indirect strategy: the properties 
satisfied by the supermultiplets from general principles determines them 
uniquely. The precise steps are discussed in the following.

\paragraph{Global symmetries:} In section~\ref{sec:global} the 
perturbations of black holes in the dilute gas regime were classified 
under the global $SU(2)\times USp(6)$ symmetry. This symmetry is preserved 
by the near horizon geometry; so there can be no mixing between the 
hyper-multiplet, the vector-multiplet, the massive gravitino multiplet, 
and the $N=2$ graviton multiplet. The field content of these multiplets 
was written out in the end of the previous section.

\paragraph{Superconformal symmetry:}
The near horizon geometry is $AdS_3\times S^2$~\cite{skenderis97a}.
This background can be expressed as the group 
manifold\footnote{Global properties of the groups are disregarded in 
the present work, {\it e.g.}, $SO(3)\simeq SU(2)$.}:
\be
AdS_3\times S^2
\simeq {SL(2,\IR)_L\times SL(2,\IR)_R 
\over SL(2,\IR)_{\rm diag}} \times {SU(2)\over U(1)}~.
\ee
In this form the bosonic symmetries are manifest. The 
$SO(2,2)\simeq SL(2,\IR)_L\times SL(2,\IR)_R$ is the conformal group,
with quantum numbers $(h,{\bar h})$; and the $SL(2,\IR)_R\times SU(2)$ 
symmetry is the bosonic subgroup of the supergroup $SU(2|1,1)_R$.
The background preserves the supersymmetry, and so its perturbations 
form representations of $SU(2|1,1)_R$~\cite{guna86}:
\bea
\begin{array}{ccc}
 {\bar h} &  {\bar j} & SU(2)\\
k+{1\over 2} & k+{1\over 2} & {\bf 1}   \\ 
k+1          & k            & {\bf 2}   \\ 
k+{3\over 2} & k-{1\over 2} & {\bf 1}   \\ 
\end{array}
\label{eq:scm}
\eea
where $k$ can be integer or half-integer. These multiplets are perhaps 
more familiar as the short representations of $N=4$ supersymmetry in 
two dimensions. The numbers of fermions and bosons at each level coincide
and equals $2(2k+1)$. Note that the $SL(2,\IR)_L$ is not a subgroup of 
any supergroup, so every element 
of the superconformal multiplet has identical eigenvalue $h$ of the 
$SL(2,\IR)_L$ generator $L_0$. 

\paragraph{Light cone helicities:}
It is simplest to analyze the physical spectrum in the light-cone 
frame, where all modes are physical. The local $SO(4,1)$ Lorentz 
group\footnote{We emphasize that this is not an isometry of the manifold, 
but it is realized locally in the tangent space.} has little group $SO(3)$ 
for massless fields; so the possible light-cone helicities for fields with 
spin $s$ are:
\be
\lambda=-s,-s+1,\cdots,s.
\ee
The light-cone can be chosen in the $t-x^{11}$ plane of the $AdS_3$,
where the $11$th dimension is along the intersection of the $M5$-branes,
and then the helicity is also given by $\lambda = h - {\bar h}$.
Thus, the transformation properties under the $D=5$ Lorentz group
give the possible values of the $AdS_3$ spin
$s_0=h-{\bar h}$, and their degeneracies.

\paragraph{The leading eigenvalues:}
Given a superconformal multiplet eq.~\ref{eq:scm} with parameter $k$,
all the higher level multiplets $k+1,k+2,\cdots $ correspond to the 
expansion in partial waves\footnote{From the algebraic point of 
view the partial waves are generated by the $SU(2)$ level $1$ currents 
of the $N=4$ superconformal algebra.}. However, the {\it smallest} value 
of $k$ must be determined from separate considerations. The first step
is to expand the fields in their $D=4$ components and exploit
that the angular momentum of a $D=4$ field with spin $s$ satisfies
${\bar j}\geq s$ so that the tower of states is precisely 
${\bar j}=s,s+1,\cdots $. This observation gives the complete set of 
angular momenta for the fields in a given representation of $USp(6)$.
However, the relation between these towers and specific spacetime fields
is not clear, because different fields with the same ${\bar j}$ can mix. 

We will resolve the remaining ambiguity by requiring that the $USp(6)$ 
multiplets can be reassembled into $SU(6)$ multiplets. This is a
definite requirement for extreme configurations, including the state
with three intersecting $M5$-branes and no momentum along the
common string\footnote{There is an ordering ambiguity in the problem:
we approach the near horizon limit first; and then take the momentum 
along the string to zero. Additionally, the leading eigenvalues of the 
super-multiplets could be interchanged in the dilute gas limit, where 
the symmetry of the central charge matrix is only $USp(6)$.}.


\vspace{0.2in}
Altogether these observations give a practical way to find all the 
superconformal multiplets: consider fields in a given $USp(6)$ 
representation and use their $D=5$ Lorentz-properties to find the 
possible helicities and their degeneracies; now simply choose 
superconformal multiplets in the unique way that exhausts all the values 
of the helicity. This can be done systematically, noting that each 
element in the superconformal multiplet has the same value of $h$, so 
the highest $SU(2)$ component, given in the first row of eq.~\ref{eq:scm}, 
is the entry with the minimal $AdS_3$ spin $s_0=h-{\bar h}$. After
the form of the multiplets has been found, the leading eigenvalues are 
determined by expansion in $D=4$ components, as described above.

In the following we make this procedure explicit by considering each 
of the $N=2$, $D=5$ multiplets in turn.

\paragraph{Hyper-multiplet:}   
There are only scalars and spin-${1\over 2}$ fermions, and all 
components fit in a single superconformal multiplet:
\bea
\begin{array}{ccc}
(h, {\bar h}) &  {\bar j} & SU(2)\times USp(6) \\
(l+1,l+{1\over 2}) & l+{1\over 2} & ({\bf 1},{\bf 14^\prime} )   \\ 
(l+1,l+1)          & l            & ({\bf 2},{\bf 14^\prime} )   \\ 
(l+1,l+{3\over 2}) & l-{1\over 2} & ({\bf 1},{\bf 14^\prime} )   \\ 
\end{array}
\label{eq:hyper}
\eea
The required partial wave numbers for scalars and fermions 
in four dimensions determine that $l=0,1,\cdots$. We use 
the convention that multiplets with ${\bar j}<0$ are absent. In other
words, the $l=0$ entry is a particularly short multiplet, where the
vacuum is acted on nontrivially by only one of the four supercharges. 

The entry in the centre row are the minimally coupled scalars whose 
conformal weights have previously been identified by explicit 
calculation~\cite{greybody2,gpartial,mpartial}; 
the results agree. By supersymmetry, the fermion with $l=0$ should
share the property with minimally coupled scalars that, for each value 
of the helicity, the absorption cross-section 
$\sigma_{\rm abs}(\omega\rightarrow 0) = A_4$, where $A_4$ is the area 
of the black hole; in contrast, $\sigma_{\rm abs}(\omega\rightarrow 0) = 0$ 
for fermions that satisfy the Weyl equation~\cite{gibbons96,cl97d}. 
It would be interesting to verify this prediction by explicit calculation.

\paragraph{Vector multiplets:}   
The helicities of the $D=5$ vectors are $h-{\bar h}=\pm 1,0$; and the
scalar has  $h-{\bar h}=0$. This determines the supermultiplets as:
\bea
\begin{array}{ccc}
( h , {\bar h}) &  {\bar j} & SU(2)\times USp(6) \\
(l+1,l+1) & l+1  & ({\bf 1},{\bf 14})                  \\ 
(l+1,l+{3\over 2}) & l+{1\over 2} & ({\bf 2},{\bf 14})   \\ 
(l+1,l+2) & l & ({\bf 1},{\bf 14})   \\ 
\end{array}
\label{eq:vector1}
\eea
and
\bea
\begin{array}{ccc}
( h , {\bar h}) &  {\bar j} & SU(2)\times USp(6) \\
(l+2,l+1) & l+1  & ({\bf 1},{\bf 14})                  \\ 
(l+2,l+{3\over 2}) & l+{1\over 2} & ({\bf 2},{\bf 14})   \\ 
(l+2,l+2) & l & ({\bf 1},{\bf 14})   \\ 
\end{array}
\label{eq:vector2}
\eea
The helicities of the fermions work out too, as they should. From the 
$D=4$ perspective there are two scalars, so there must be
exactly two towers with the angular momenta ${\bar j}=0,1,\cdots$ ;
this requires that both multiplets have $l=0,1,\cdots$.

The $AdS_3$ scalars, with spin $s_0=h-{\bar h}=0$, are in general 
mixtures of the $D=5$ vectors and the $D=5$ fixed scalars.
However, there are precisely two states with $j=0$, and the one
with $s_0=-1$ must be part of the $D=5$ vector; so the state with $s_0=j=0$ 
can include no vector contribution. Thus it must be the fixed scalar in 
the S-wave, with no mixing. The conformal weights $(h,{\bar h})=(2,2)$ of 
this state agree with those that have previously been identified by 
direct calculation~\cite{cgkt,krasnitz97}.

\paragraph{Massive gravitino multiplets:}   
In this case the $D=5$ vector transforms in the ${\bf 2}$ of the
global $SU(2)$; so its helicities determines the three supermultiplets:
\bea
\begin{array}{ccc}
( h , {\bar h}) &  {\bar j} & SU(2)\times USp(6) \\
(l,l+{1\over 2}) & l+{1\over 2} & ({\bf 1},{\bf 6})   \\ 
(l,l+1) & l  &                    ({\bf 2},{\bf 6})   \\ 
(l,l+{3\over 2}) & l-{1\over 2} & ({\bf 1},{\bf 6})   \\ 
 & & \\
(l+1,l+{1\over 2}) & l+{1\over 2} & ({\bf 1},{\bf 6})   \\ 
(l+1,l+1) & l  &                    ({\bf 2},{\bf 6})   \\ 
(l+1,l+{3\over 2}) & l-{1\over 2} & ({\bf 1},{\bf 6})   \\ 
 & & \\
(l+2,l+{1\over 2}) & l+{1\over 2} & ({\bf 1},{\bf 6})   \\ 
(l+2,l+1) & l  &                    ({\bf 2},{\bf 6})   \\ 
(l+2,l+{3\over 2}) & l-{1\over 2} & ({\bf 1},{\bf 6})   \\ 
\end{array}
\eea
Again, the fermion helicities serve as a check. 

The $D=5$ vector decomposes into a $D=4$ vector, and a $D=4$ scalar.
Since ${\bar j}=l$ for the bosons in these tables it follows that 
exactly two of the tables have $l=1,2,\cdots $, and one has 
$l=0,1,\cdots $. 

The middle table is identical to the result for the hyper-multiplet
eq.~\ref{eq:hyper}, except for the $USp(6)$ representation. Therefore
the condition that the states can be assembled into $SU(6)$ multiplets 
forces that the middle table has range $l=0,1,\cdots $, just as the 
hyper-multiplet. Accordingly, the other two representations have
$l=1,2,\cdots $.


\paragraph{Graviton multiplet:}   
Finally, the $D=5$ graviton has light cone helicities 
$s_0=h-{\bar h}=\pm 2,\pm 1,0$ ; so, together with the 
$D=5$ vector helicities $s_0=h-{\bar h}=\pm 1,0$, 
the supermultiplets are determined as:
\bea
\begin{array}{ccc}
( h , {\bar h}) &  {\bar j} & SU(2)\times USp(6) \\
(l,l+1) & l+1  & ({\bf 1},{\bf 1})   \\ 
(l,l+{3\over 2}) & l+{1\over 2} & ({\bf 2},{\bf 1})   \\ 
(l,l+2) & l  & ({\bf 1},{\bf 1})   \\ 
 && \\
(l+1,l+1) & l+1  & ({\bf 1},{\bf 1})   \\ 
(l+1,l+{3\over 2}) & l+{1\over 2} & ({\bf 2},{\bf 1})   \\ 
(l+1,l+2) & l  & ({\bf 1},{\bf 1})   \\ 
 && \\
(l+2,l+1) & l+1  & ({\bf 1},{\bf 1})   \\ 
(l+2,l+{3\over 2}) & l+{1\over 2} & ({\bf 2},{\bf 1})   \\ 
(l+2,l+2) & l  & ({\bf 1},{\bf 1})   \\ 
 && \\
(l+3,l+1) & l+1  & ({\bf 1},{\bf 1})   \\ 
(l+3,l+{3\over 2}) & l+{1\over 2} & ({\bf 2},{\bf 1})   \\ 
(l+3,l+2) & l  & ({\bf 1},{\bf 1})   \\ 
\end{array}
\eea
 From the $D=4$ perspective the boson fields are $2$ scalars, $2$
vectors, and $1$ graviton. The vectors and the graviton each have
two degrees of freedom and the scalar has one. Thus there must be
two towers with $SU(2)$ quantum numbers ${\bar j}=0,1,\cdots $,
four with ${\bar j}=1,2,\cdots $ and two with
${\bar j}=2,3,\cdots $. Comparing with the tables we find that
two supermultiplets have indices $l=0,1,\cdots $, and two have 
$l=1,2,\cdots $. 

Two graviton multiplets must combine with the vector multiplets 
eqs.~\ref{eq:vector1}-\ref{eq:vector2} and form larger multiplets, with 
the global group $SU(6)$. This condition determines that it is the
middle two multiplets that have $l=0,1,\cdots $; the first and
the last multiplet has $l=1,2,\cdots $


\vspace{0.2in}
The wave functions of particles with spin in the background of
dilute gas black holes have previously been considered 
in~\cite{gubser,hosomichi,cl97d}. It was found that the leading 
Hawking emission of particles with spin $s$ at low frequency is 
controlled by the conformal weights $(1,1+s)$~\cite{cl97d}. This result 
was based on ``minimal'' couplings, {\it i.e.} only the coupling to 
gravity was taken into account. In the present work, supergravity
specifies additional background fields. Nevertheless, 
the tables do have entries with the leading quantum numbers $(1,1+s)$. 
This indicates the interpretation of Hawking emission of particles with 
spin remains qualitatively correct, even though a complete analysis of 
the greybody factors for the $N=8$ black holes has not yet been carried 
out.
 
\subsection{The Chiral Primary Fields}
\label{sec:chiral4d}
The chiral primaries are useful because they generate the complete 
supermultiplet. They are defined as the states that satisfy 
${\bar h}={\bar j}$. More precisely, the representation ${\bar j}$ of 
the $SU(2)$ rotation group has elements with projections on some axis 
${\bar j}_3 = -{\bar j},\cdots, {\bar j}$. The states with 
${\bar h}={\bar j}_3$ are the chiral primaries; those with 
${\bar h}=-{\bar j}_3$ are anti-chiral primaries. We will informally refer
to all these fields as chiral primaries.

The first entry of each table in the previous section corresponds to a 
chiral primary. In the extreme limit the fields form representations 
of $SU(6)$, and we collect them as:
\bea
\begin{array}{cccc}
 h & {\bar h}={\bar j_3} & SU(6) & l\\
l+1&l+{1\over 2} & {\bf 20} & 0,1,\cdots \\
l+1&l+1  & {\bf 15} & 0,1,\cdots  \\  
l+2&l+1  & {\bf 15} & 0,1,\cdots  \\ 
l  &l+{1\over 2} & {\bf 6}  & 1,2,\cdots \\ 
l+2&l+{1\over 2} & {\bf 6}  & 1,2,\cdots \\ 
l  & l+1  & {\bf 1}  & 1,2,\cdots \\ 
l+3& l+1  & {\bf 1}  & 1,2,\cdots \\ 
\end{array}
\label{eq:4dspec}
\eea
All the chiral primaries are singlets under the global $SU(2)$.

The $AdS/CFT$ correspondance predicts that this table gives the complete 
list of chiral primaries in the effective string theory describing 
three orthogonally intersecting $M5$-branes. In supergravity, it gives 
all the conformal weights underlying the greybody factors of extreme
black holes. The result for the dilute gas regime is found by 
decomposing the $SU(6)$ representation into $USp(6)$, according to
eqs.~\ref{eq:15r}-\ref{eq:20r}.

The physical significance of the precise ranges of $l$ is not clear. It 
is known that the ``missing'' states with $l=0$ generally correspond to 
modes that are pure gauge. However, it is possible such modes induce 
boundary states at AdS-infinity, and so they may nevertheless play a
role in the $AdS/CFT$ correspondance. The discussion in the following 
paragraphs disregard the range of $l$. 

In the dilute gas regime, all the chiral primaries at a given level 
can be generated from the ${\bf 14}^\prime$ state, by acting with two 
operators of the form ${\cal Q}^A_{h,{\bar h}}$, where $A$ is a 
$USp(6)$ vector 
index, ${\bar h}={1\over 2}$, and $h$ takes the two values $h=0$ and $h=1$, 
respectively. This suggests that these operators, together with their 
complex conjugates, are symmetries of the chiral algebra. The 
${\cal Q}^A_{h,{\bar h}}$ have half-integer spin; so this global
symmetry is itself a supersymmetry. Global symmetries are generally
robust; thus it is reasonable to expect that the ${\cal Q}^A_{h,{\bar h}}$
persist in the full interacting string theory. 

The ${\cal Q}^A_{h,{\bar h}}$ admit a spacetime interpretation
in terms of the {\it broken supersymmetry}. Namely, the infinitesimal
generators of the broken supersymmetries act nontrivially on the
background creating fermionic ``perturbations''. Repeated actions 
form a closed algebra, by the underlying supersymmetry. The restriction 
of this algebra of broken supersymmetry to the chiral operators gives the 
${\cal Q}^A_{h,{\bar h}}$. These operators must have definite
transformation properties under the preserved symmetries. They are
in the ${\bf 6}$ of $USp(6)$, by restriction of the original ${\bf 8}$ 
of $SU(8)$. Generators of supersymmetry are spin-${1\over 2}$ so
${\bar h}={1\over 2}$, to preserve the chiral condition. Finally, there
are two $D=5$ helicities, so $h-{\bar h}=\pm {1\over 2}$ determines
the two values of the conformal weight $h$ as $h=0$ and $h=1$.  
It would be interesting to investigate the properties of the 
${\cal Q}^A_{h,{\bar h}}$ in more detail.

\section{$N=8$ Black Holes in Five Dimensions}
\label{sec:5d}
The perturbation spectrum of five-dimensional black holes and their 
associated six-dimensional black strings has been considered recently 
by several workers~\cite{strom98a,martinecads3,sezginads3}. In particular, 
the linearized equations of motion have been decoupled completely
by explicit calculation, in the case of $N=4$ 
supergravity~\cite{strom98a,sezginads3}. The purpose of the present 
section is to extend this result to $N=8$ supergravity. In other words, 
we consider the background $AdS_3\times S^3\times M$, with the
small manifold $M=T^4$, rather than $M=K3$.

The matter content of $N=8$ supergravity differs from the
matter content of $N=4$ supergravity by having $n=5$ anti-selfdual 
tensor supermultiplets, instead of $n=21$. There are also additional
fields, namely $4$ gravitini, $16$ vector fields and $20$ fermions, 
with counting from the $D=6$ perspective. In the linearized approximation, 
there are no couplings between the original $N=4$ and the 
additional $N=8$ perturbations\footnote{It is consistent to take the 
$16$ vector fields vanishing in the background, so they can appear 
only in quadratic order in the variations with respect to any of the 
$N=4$ fields. Alternatively, the decoupling follows from global 
symmetries.}; so the results obtained previously for $N=4$ remain valid,
except that now $n=5$. However, the additional gravitino multiplet
requires further considerations.

The conformal weights of the additional multiplet can be worked out using 
group theory, as in the previous section. A new feature is that,
for massless fields, the little group $SO(4)\simeq SU(2)\times SU(2)$ 
of the Lorentz group $SO(5,1)$ is not simple; so there are two independent 
helicities $\lambda$ and 
${\bar\lambda}$. The $AdS_3$ spin is given in terms of the helicities as 
$s_0=h-{\bar h}=\lambda+{\bar\lambda}$. The result obtained from
group theory agrees with the one given below, in 
eq.~\ref{eq:vectorrep}. However, in the present section 
we follow the explicit calculation of~\cite{sezginads3}, and decouple 
the linearized equations of motion explicitly for the bosons.

\subsection{Decoupling of the Equations of Motion}

The linearized equations of motion for the vector fields in $D=6$ SUGRA
is:
\be
\nabla^I F_{IJ}^\alpha - {1\over 6}\epsilon^{~KLMNO}_{J}~
(\Gamma_A)^{\alpha}_{~\beta}F^\beta_{KL}~H_{MNO}^A = 0~.
\ee
These wave equations follow from duality, with the numerical coefficient 
determined by explicit dimensional reduction from $D=11$ of selected 
components. The duality group $SO(5,5)$ has 
spinor indices $\alpha,\beta= 1,\cdots,16$ and vector index
$A = 1,\cdots,10$. The $SO(5,1)$ Lorentz group has 
vector indices $I,J,\cdots = 0,\cdots,5$ that decompose into 
the $AdS_3$ indices $\mu,\nu\cdots = 0,1,2$ and the $S^3$ indices
$a,b,\cdots = 3,4,5$. 

The only nonvanishing matter field in the black hole
background is a selfdual component of the antisymmetric tensor, 
decomposed as:
\be
H^5_{abc} = \epsilon_{abc}~~~;~H_{\mu\nu\rho}^5=\epsilon_{\mu\nu\rho}~,
\label{eq:hback}
\ee
in units where the cosmological constant is $\Lambda = -l^2 = -1$.

The vacuum breaks the duality group $SO(5,5)\rightarrow SO(5)\times SO(5)$,
and the black hole background breaks this global symmetry further as
$SO(5)\times SO(5)\rightarrow SO(4)\times SO(5)$. The field 
strength transforms as the spinor ${\bf 16}$ under $SO(5,5)$, and thus 
as the bispinor $({\bf 4},{\bf 4})$ in the $SO(5)\times SO(5)$ 
symmetric vacuum. In the $SO(4)\times SO(5)$ symmetric background 
created by the black hole this representation is further broken to 
$({\bf 2_+},{\bf 4})\oplus ({\bf 2_-},{\bf 4})$. 
The Gamma-matrices $(\Gamma_A)^{\alpha}_{~\beta}$ of $SO(5,5)$
can be decomposed according to this symmetry breaking pattern. Then
the two representations are distinguished by the eigenvalue 
$P=\pm 1$ of:
\be
(\Gamma_{11})^{\alpha}_{~\beta}F^\beta_{KL} = P F^\alpha_{KL}~.
\label{eq:polar}
\ee
For definiteness we concentrate on $P=1$ and so the representation
$({\bf 2_+},{\bf 4})$. The result for the alternative projection 
$P=-1$ will be recovered in due course. Thus the equation of motion becomes:
\be
\nabla^I F_{IJ} - {1\over 6}\epsilon^{~KLMNO}_{J}
F_{KL} H_{MNO}^5  = 0~,
\ee
or, in view of eq.~\ref{eq:hback}~\footnote{We use conventions
where $\epsilon^{\mu\nu\rho abc}=-\epsilon^{abc \mu\nu\rho}
=\epsilon^{\mu\nu\rho}\epsilon^{abc}$.}:
\bea
\nabla^I F_{Ia}- 
\epsilon^{~bc}_a F_{bc} &=& 0~,
\label{eq:ads3scalar}
\\
\nabla^I F_{I\mu} - \epsilon^{~\nu\rho}_\mu 
F_{\nu\rho} &=& 0~.
\label{eq:ads3vector}
\eea

The general expansion in spherical harmonics on $S^3$ is:
\bea
A_\mu (x,y) &=& \sum_l A^{(l,0)}_\mu (x) Y^{(l,0)}(y)~, \\
A_a (x,y) &=& \sum_l [A^{(l,\pm 1)} (x) Y_a^{(l,\pm 1)}(y)
+  A^{(l,0)} (x)\partial_a Y^{(l,0)}(y)]~,
\label{eq:aax}
\eea
where the coordinates on $AdS_3$ and $S^3$ are denoted $x$ and $y$,
respectively. The spherical harmonics satisfy~\cite{sezginads3}:
\bea
\nabla^2_y Y^{(l,0)} &=& [ 1- (l+1)^2 ] Y^{(l,0)}~, \\
\nabla^2_y Y^{(l,\pm 1)}_a &=& [ 2- (l+1)^2 ] Y^{(l,\pm 1)}_a~,\\
\nabla^a Y_a ^{(l,\pm 1)} &=& 0~,\\
\epsilon_a^{~bc}\partial_b Y_c ^{(l,\pm 1)} &=& 
\pm (l+1)Y^{(l,\pm 1)}_a~.
\label{eq:yrel}
\eea
It is straightforward to show that a gauge transformation can be chosen
so that:
\be
\partial^a A_a (x,y) = 0~.
\label{eq:ldd}
\ee
This condition defines the Lorentz-DeDonder gauge. In this gauge 
$A^{(l,0)}=0$ so that the last term in eq.~\ref{eq:aax} vanishes.
The gauge condition eq.~\ref{eq:ldd} allows further gauge transformations
of the form:
\be
\delta A_\mu^{(0,0)} (x) = \partial_\mu \Lambda (x)~.
\label{eq:resgauge}
\ee
These are generated by the 0-mode $Y^{(0,0)}$ on the sphere.

The equations of motion eq.~\ref{eq:ads3scalar} for the
$AdS_3$ scalars become:
\be
\sum_l [\nabla^\mu \partial_\mu - (l+1)^2\mp 2(l+1)] A^{(l,\pm 1)}
Y_a^{(l,\pm 1)}
-\sum_l \nabla^\mu A^{(l,0)}_\mu\partial_a Y^{(l,0)} = 0~,
\ee
in the Lorentz-DeDonder gauge. 
Orthogonality relations of the spherical harmonics gives:
\be
\nabla^\mu A^{(l,0)}_\mu (x)= 0~,
\ee
so that the longitudinal modes decouple, as they should. The remaining 
equations are those of minimally coupled scalars in the $AdS_3$, with 
effective masses:
\be
m^2 = (l+1)^2 \pm 2(l+1)~.
\ee
Then the conformal weights given through:
\be
m^2 = 4h ( h-1 )~,
\ee
become:
\be
h = {l+2\pm 1\over 2}~.
\ee
The $SU(2)$ quantum numbers of the fields are related to the indices
$(l_1,l_2)$ of the spherical harmonics through 
$(j,{\bar j})= ({l_1+l_2\over 2},{l_1-l_2\over 2})$.
Moreover, the fields are $AdS_3$ scalars so $h={\bar h}$. Thus
we arrive at the following table:
\bea
\begin{array}{cccc}
(h,{\bar h}) & (j,{\bar j} )& SO(4)\times SO(5) & l \\
({l+3\over 2},{l+3\over 2}) & ({l+1\over 2}, {l-1\over 2} ) &   
(2_{+},4) & 1,2,\cdots  \\
({l+1\over 2},{l+1\over 2}) & ({l-1\over 2}, {l+1\over 2} ) & 
(2_{+},4) & 1,2,\cdots \\
\end{array}
\label{eq:scalartable}
\eea

Next we consider the $AdS_3$ vectors. Their equation of motion is
eq.~\ref{eq:ads3vector}. In the Lorentz-DeDonder gauge 
eq.~\ref{eq:ldd} we immidiately find independent equations for each 
partial wave:
\be
\nabla^\nu \partial_{[\nu }A^{(l,0)}_{\mu ]}
-\epsilon_\mu^{~\nu\rho} \partial_{[\nu }A^{(l,0)}_{\rho ]}
= l(l+2)A^{(l,0)}_{\mu}~.
\label{eq:veceom}
\ee
It is simplest to analyze this equation by taking advantage of the 
$SL(2,\IR)_L\times SL(2,\IR)_R$ symmetry. From this point 
of view it is clear that vector fields in $AdS_3$ satisfy 
$h-{\bar h}=\pm 1$ and we can view eq.~\ref{eq:veceom} as an equation 
for the energy $E_0 = h + {\bar h}$. The $SL(2,\IR)$ representation 
theory is analogous to that of $SU(2)$. We need the equation analogous
to eq.~\ref{eq:yrel}, namely:
\be
\epsilon_{\mu}^{~\nu\rho}\partial_\nu A_\rho^{(l,0)}
= \mp (h + {\bar h} -1)A_\mu^{(l,0)}~.
\label{eq:ahh}
\ee
We find:
\be
(h + {\bar h} -1)^2 \pm 2(h + {\bar h} -1) = l( l+2)~,
\ee
so
$h+{\bar h} = l+1$ and $h+{\bar h} = l+3$ for
$ h-{\bar h} =\pm 1 $, respectively. We thus arrive at the 
table:
\bea
\begin{array}{cccc}
(h,{\bar h}) & (j,{\bar j} )& SO(4)\times SO(5) & l \\
({l+2\over 2},{l\over 2}) & ({l\over 2}, {l\over 2} ) & 
(2_{+},4) & 1,2,\cdots \\
({l+2\over 2},{l+4\over 2}) & ({l\over 2}, {l\over 2} ) &   
(2_{+},4) & 0,1,2,\cdots  \\
\end{array}
\label{eq:vectortable}
\eea
Note that the functions $Y^{(l,0)}$ are the standard 
scalar spherical harmonics with indices $l=0,1\cdots$. However, 
according to eq.~\ref{eq:ahh} the corresponding mode is constant 
when $h+{\bar h} = 1$, and thus this mode is not propagating. This
absence of the $l=0$ mode is related to the residual gauge transformation 
eq.~\ref{eq:resgauge}.

At this point we should recall that we have omitted half the story,
due to the choice of $P=+1$ after eq.~\ref{eq:polar}. Accordingly,
tables~\ref{eq:scalartable} and~\ref{eq:vectortable} summarizing 
the results contain only entries with $SO(4)$ quantum numbers $2_+$. 
The fields with $P=-1$ and thus the global quantum numbers $(2_-,4)$, 
are the complex conjugates of those we have considered. They correspond 
to the entries of table~\ref{eq:scalartable} with 
$j\leftrightarrow {\bar j}$, and the entries of
table~\ref{eq:vectortable} 
with $h\leftrightarrow {\bar h}$.

\subsection{Superconformal Multiplets}
We shift the index $l$ of the second entry in both tables
(~\ref{eq:vectortable} and~\ref{eq:scalartable}) so that 
$l=0,1,\cdots$; and, recalling the convention 
that multiplets with $SU(2)$ indices $j=-1$ or ${\bar j}=-1$ vanish, we
allow $l=0$ for the first entry of eq.~\ref{eq:scalartable}. 
Then all the modes in the supermultiplet can be assembled into the table:
\bea
\begin{array}{ccc}
(h,{\bar h}) & (j,{\bar j} )& SO(4)\times SO(5) \\
({l+2\over 2},{l+2\over 2}) & ({l+2\over 2}, {l\over 2} ) &   (2_{-},4) \\
({l+3\over 2},{l+3\over 2}) & ({l+1\over 2}, {l-1\over 2} ) & (2_{+},4) \\
({l+2\over 2},{l+1\over 2}) & ({l+2\over 2}, {l+1\over 2} ) & (1,4) \\
({l+2\over 2},{l+3\over 2}) & ({l+2\over 2}, {l-1\over 2} ) & (1,4) \\
({l+3\over 2},{l+2\over 2}) & ({l+1\over 2}, {l\over 2} ) &   (4,4) \\
({l+4\over 2},{l+3\over 2}) & ({l\over 2}, {l-1\over 2} ) &   (1,4) \\
({l+3\over 2},{l+1\over 2}) & ({l+1\over 2}, {l+1\over 2} ) & (2_{+},4) \\
({l+4\over 2},{l+2\over 2}) & ({l\over 2}, {l\over 2} ) &     (2_{-},4) \\
({l+4\over 2},{l+1\over 2}) & ({l\over 2}, {l+1\over 2} ) &   (1,4) \\
\end{array}
\label{eq:vectorrep}
\eea
plus its complex conjugate. The fermionic entries were inferred 
by supersymmetry, discussed below. At each level $l$ there are 
$8(l+1)(l+2)$ bosons and $8(l+1)(l+2)$ fermions.

The underlying symmetry structure of the near horizon geometry
of black holes in five dimensions is the factorized supergroup 
$SU(2|1,1)_L\times SU(2|1,1)_R$. In particular, the two-dimensional 
supersymmetry is $(4,4)$. It follows that the states can be organized 
into supermultiplets under left and right moving generators independently. 
Indeed, the table above is the tensor product of two $N=4$ multiplets of 
the form given in eq.~\ref{eq:scm}, with parameters $k_L=(l+1)/2$ and 
$k_R=l/2$; and the complex conjugate multiplet similarly derives from 
$k_L=l/2$ and $k_R=(l+1)/2$. The chiralities of the $D=6$ vectors, 
and the quantum numbers of the fermions were in fact determined
precisely by demanding this structure.

The chiral primary fields with respect to any $(2,2)$ subalgebra of the 
$(4,4)$ supersymmetry satisfy $(h,{\bar h})=(j,{\bar j})$~\footnote{More 
precisely the $(j,{\bar j})$ representation of the $SU(2)\times SU(2)$
has components with $j_3 = -j,\cdots, j$ and 
${\bar j}_3 = -{\bar j},\cdots, {\bar j}$. The chiral fields are
those components that satisfy $(h,{\bar h})=(\pm j_3,\pm{\bar j_3})$, 
with the four possible signs corresponding to elements in the $(c,c)$-, 
$(c,a)$-, $(a,c)$- and $(a,a)$-rings.}. In the table above, the chiral 
primary fields are the entries with 
$(h,{\bar h}) = ({l+2\over 2}, {l+1\over 2})$. 
Similar tables, for the fields present in the $N=4$ case, 
are given in~\cite{sezginads3}. This allows the assembly of a complete 
list of chiral primaries in the $N=8$ black hole background. It is:
\bea
\begin{array}{cc}
( h , {\bar h}) & SO(4)\times SO(5) \\
({l+3\over 2},{l+1\over 2})+{\rm c.c.} & (1,1)   \\ 
({l+1\over 2},{l+1\over 2}) & (1,5) \\ 
({l+2\over 2},{l+2\over 2}) & (1,1) \\
({l+2\over 2},{l+1\over 2})+{\rm c.c.} & (1,4)   \\ 
\end{array}
\label{eq:chirals}
\eea
where $l=0,1,\cdots$. This representation of the complete perturbation 
spectrum is convenient for comparison with string theory.

\section{Microscopic Interpretation}
\label{sec:string}
According to Maldacena's conjecture~\cite{juanads} the spectrum of black 
hole perturbations is identical to that of the underlying string theory. 
For five dimensional black holes the string theory result is given in the
work of Strominger and Vafa~\cite{strom96a}. This sets the stage for a 
detailed comparison, elaborating the one given 
in~\cite{strom98a,martinecads3}.

Consider the string theory spectrum of $n_1$ $D1$-branes and 
$n_5$ $D5$-branes, wrapped on a small Calabi-Yau manifold $M$ 
with two complex dimensions, {\it i.e.} $M=T^4$ or $M=K3$. 
The spectrum is given by a superconformal $\sigma$-model on the target 
space~\cite{strom96a}:
\be
{\cal C} = M^k /\Sigma_k
\label{eq:symorb}
\ee
where the level $k=n_1 n_5$. The RR states of the $\sigma$-model
are in one-to-one correspondance with the elements of the cohomology of 
the target space. It is a property of the symmetric orbifold construction
that the cohomology of ${\cal C}$ can be constructed as a Fock space
over the cohomology of $M$~\cite{vw94,dmvv}. In particular, the level 
$n$ $(p,q)$-form of $H^{*}({\cal C})$, denoted $(p,q)_n$, is the permutation 
invariant product of $n$ $(p,q)$-forms of $H^{*}(M)$. More general 
elements in  $H^{*}({\cal C})$ are generated by combining these 
elementary ones, as in the construction of a Fock space.

We are interested in the chiral operators, more precisely the $(c,c)$ 
chiral ring. These are in the NS-NS sector, related to the 
RR-sector by spectral flow. Their spectrum is~\cite{strom98a}:
\be
(h,{\bar h})=(j_3,{\bar j}_3) = {1\over 2}\sum_i (p_i-1+n_i, q_i-1+n_i)~.
\ee 
The chiral operators are organized in a Fock-space whose building 
blocks are the ``single particle operators'', just as the RR-states.
The conformal weights of the single particle chiral primaries are given 
by the individual terms in the expression above, with levels
$n=1,2,\cdots$, except that $n=2,3,\cdots$ for $p=q=0$ because 
$(0,0)_1$ corresponds to the standard $NS-NS$ vacuum. Thus, their 
degeneracies are:
\bea
\begin{array}{cc}
( h , {\bar h}) & {\rm degeneracy}\\
({l+2\over 2},{l\over 2})+{\rm c.c.} & h_{2,0}+{\rm c.c.} \\ 
({l+1\over 2},{l+1\over 2}) & h_{0,0}+h_{1,1} \\ 
({l+2\over 2},{l+2\over 2}) &  h_{2,2}\\
({l+1\over 2},{l\over 2})+{\rm c.c.} & h_{1,0}+{\rm c.c.} \\ 
({l+2\over 2},{l+1\over 2})+{\rm c.c.} & h_{2,1}+{\rm c.c.} \\ 
\end{array}
\label{eq:chiral}
\eea
where $l=0,1,\cdots$.

The spectrum of chiral operators in string theory should be compared with 
that of black hole perturbations. In the linearized approximation, the 
perturbations can be superimposed, and therefore they form a Fock space, 
just as in string theory. Thus, it is sufficient to compare the 
{\it single particle} operators on the two sides. Moreover, it is 
only the chiral primaries that are needed, since these operators generate 
the complete supermultiplet. 

For $T^4$ the ``scalar'' Betti-numbers are $h_{2,2}=1$ and 
$h_{0,0}+h_{1,1}=5$. This gives perfect agreement between string 
theory (table~\ref{eq:chiral}) and supergravity (table~\ref{eq:chirals}),
as in~\cite{strom98a} (except that there $M=K3$). The remaining
Betti-numbers are $h_{2,0}=1$ and $h_{2,1}=h_{1,0}=2$; and their complex 
conjugates. Here the degeneracies of string theory and supergravity
agree again, but there is a minor discrepancy in the conformal
weights: the first element of the $(2,0)$ and the $(1,0)$ string towers 
are absent in the supergravity description. The states that are missing 
are not propagating on the supergravity side because they are pure 
gauge modes; so it is quite proper that they are not included in the 
table. However, they may nevertheless induce physical degrees of 
freedom on the boundary at infinity; if these modes are included the 
agreement is restored. It would be interesting to work out this 
possibility in more detail.

There is also an important structural difference between string theory 
and supergravity: in string theory the levels take a finite range 
$n=1,\cdots, k$, but in the standard Fock space description they do not. 
This is the ``stringy exclusion principle''~\cite{strom98a}. Moreover, 
in multiparticle states, the {\it total} occupation number (weighted 
with respect to the level) of {\it all} varieties of excitations is 
similarly bounded by $k$. Thus the stringy exclusion principle also 
applies to particles that are not identical.

Let us conclude with a few comments on the conformal field theory 
underlying the four dimensional black holes. This theory is the effective 
two dimensional theory of three orthogonally intersecting $M5$-branes, 
wrapped on a small torus with their line of intersection kept large. 
Little is known about this theory, at least in comparison with the $D1-D5$ 
bound state. In particular, there is no complete list of chiral primaries 
derived from string theory that can be compared with the perturbation 
spectrum of the near horizon geometry, displayed in table~\ref{eq:4dspec}. 
The result given in the table can be viewed as a prediction for string 
theory.

In supergravity the spectrum of perturbations forms a Fock space, 
at the linearized level. Thus, it is inherent in the construction of 
the CFT {\it via} $AdS/CFT$ correspondance that there is a large set 
of multi-particle chiral primaries, organized in a Fock space over the 
single particle chiral primaries, as in the $D1-D5$ case. This suggests 
that, in analogy with the $D1-D5$ system, we seek a $\sigma$-model on a 
symmetric orbifold with the structure given in eq.~\ref{eq:symorb}. 
In the case of intersecting $M5$ branes we expect $k=n_1 n_2 n_3$, where 
$n_i$ is the respective number of $M5$-branes. It is not {\it a priori}
clear what the manifold $M$ is; but it should be readily determined from the 
spectrum of chiral primaries~\cite{toappear}.

\vspace{0.2in} {\bf Acknowledgments:} 
I would like to thank M. Cveti\v{c}, S. Gubser, H. Verlinde and S. Mathur,
for discussions; and R. Leigh for collaboration in the initial stages 
of this project. This work is supported in part by DOE grant 
DOE-FG02-95ER40893.


\begin{thebibliography}{10}

\bibitem{juanads}
J.~Maldacena.
\newblock The large {N} limit of superconformal field theories and
  supergravity.
\newblock hep-th/9711200.

\bibitem{wittenads}
E.~Witten.
\newblock Anti-de {S}itter space and holography.
\newblock hep-th/9802150.

\bibitem{polyakovads}
S.S. Gubser, I.R. Klebanov, and A.M. Polyakov.
\newblock Gauge theory correlators from noncritical string theory.
\newblock hep-th/9802109.

\bibitem{strom96a}
A.~Strominger and C.~Vafa.
\newblock Microscopic origin of the {B}ekenstein-{H}awking entropy.
\newblock {\em Phys. Lett. B}, 379:99--104, 1996.
\newblock hep-th/9601029.

\bibitem{dmvv}
R.~Dijkgraaf, G.~Moore, E.~Verlinde, and H.~Verlinde.
\newblock Elliptic genera of symmetric products and second quantized strings.
\newblock {\em Commun. Math. Phys.}, 185:197--209, 1997.
\newblock hep-th/9608096.

\bibitem{sezginads3}
S.~Deger, A.~Kaya, E.~Sezgin, and P.~Sundell.
\newblock Spectrum of {D = 6}, {N=4B} supergravity on ${A}d{S}_3\times {S}^3$.
\newblock hep-th/9804166.

\bibitem{strom98a}
J.~Maldacena and A.~Strominger.
\newblock ${A}d{S}_3$ black holes and a stringy exclusion principle.
\newblock hep-th/9804085.

\bibitem{martinecads3}
E.~Martinec.
\newblock Matrix models of {A}d{S} gravity.
\newblock hep-th/9804111.

\bibitem{vafaads3}
C.~Vafa.
\newblock Puzzles at large {N}.
\newblock hep-th/9804172.

\bibitem{dyon}
M.~Cveti\v{c} and D.~Youm.
\newblock Dyonic {BPS}-saturated black holes of heterotic string theory on a
  six--torus.
\newblock {\em Phys. Rev. D}, 53:584, 1996.
\newblock hep-th/9507090.

\bibitem{cystrings}
M.~Cveti\v{c} and D.~Youm.
\newblock {BPS} saturated and nonextreme states in abelian {K}aluza-{K}lein
  theory and effective {N=4} supersymmetric string vacua.
\newblock In {\em STRINGS 95: Future Perspectives in String Theory}, pages
  131--147. Los Angeles, CA, 1995.
\newblock hep-th/9508058.

\bibitem{hlm}
G.~T. Horowitz, D.~A. Lowe, and J.~M. Maldacena.
\newblock Nonextremal black hole microstates and {U}-duality.
\newblock {\em Phys. Rev. Lett.}, 77:430--433, 1996.
\newblock hep-th/9603195.

\bibitem{dvv2}
R.~Dijkgraaf, E.~Verlinde, and H.~Verlinde.
\newblock Counting dyons in {N=4} string theory.
\newblock {\em Nucl. Phys. B}, 484:543--561, 1997.
\newblock hep-th/9607026.

\bibitem{witt4d}
J.~Maldacena, A.~Strominger, and E.~Witten.
\newblock Black hole entropy in {M} theory.
\newblock {\em JHEP}, 12:002, 1997.
\newblock hep-th/9711053.

\bibitem{vafa4d}
C.~Vafa.
\newblock Black holes and {C}alabi-{Y}au threefolds.
\newblock hep-th/9711067.

\bibitem{mathur}
S.~Das and S.~Mathur.
\newblock Comparing decay rates for black holes and {D}-branes.
\newblock {\em Nucl. Phys. B}, 478:561--576, 1996.
\newblock hep-th/9606185.

\bibitem{greybody}
J.~Maldacena and A.~Strominger.
\newblock Black hole greybody factors and {D}-brane spectroscopy.
\newblock {\em Phys. Rev. D}, 55:861--870, 1996.
\newblock hep-th/9609026.

\bibitem{cgkt}
C.~G. Callan, S.~S. Gubser, I.~R. Klebanov, and A.~A. Tseytlin.
\newblock Absorption of fixed scalars and the {D}-brane approach to black
  holes.
\newblock {\em Nucl. Phys. B}, 489:65--94, 1997.
\newblock hep-th/9610172.

\bibitem{greybody2}
J.~Maldacena and A.~Strominger.
\newblock Universal low-energy dynamics for rotating black holes.
\newblock {\em Phys. Rev. D}, 56:4975--4983, 1997.
\newblock hep-th/9702015.

\bibitem{krasnitz97}
M.~Krasnitz and I.~Klebanov.
\newblock Testing effective string models of black holes with fixed scalars.
\newblock {\em Phys.Rev.D}, 56:2173--217, 1997.
\newblock hep-th/9703216.

\bibitem{intscalar}
I.~Klebanov, A.~Rajaraman, and A.~Tseytlin.
\newblock Intermediate scalars and the effective string model of black holes.
\newblock {\em Nucl.Phys.B}, 503:157--176, 1997.
\newblock hep-th/9704112.

\bibitem{mpartial}
S.~Mathur.
\newblock Absorption of angular momentum by black holes and {D}-branes.
\newblock {\em Nucl. Phys. B}, 514:204--226, 1998.
\newblock hep-th/9704156.

\bibitem{gubser}
S.~Gubser.
\newblock Absorption of photons and fermions by black holes in four-dimensions.
\newblock {\em Phys.Rev.D}, 56:7854--7868, 1997.
\newblock hep-th/9706100.

\bibitem{hosomichi}
K.~Hosomichi.
\newblock Fermion emission from five-dimensional black holes.
\newblock hep-th/9711072.

\bibitem{cl97d}
M.~Cveti\v{c} and F.~Larsen.
\newblock Greybody factors for black holes in four dimensions: Particles with
  spin.
\newblock {\em Phys. Rev. D}, 57:6297--6310, 1998.
\newblock hep-th/9712118.

\bibitem{cl98a}
M.~Cveti\v{c} and F.~Larsen.
\newblock Near horizon geometry of rotating black holes in five dimensions.
\newblock hep-th/9805097.

\bibitem{deboer}
J.~DeBoer.
\newblock Talk at the {ITP}, santa barbara.
\newblock 5.7.98.

\bibitem{btzentropy}
A.~Strominger.
\newblock Black hole entropy from near horizon microstates.
\newblock {\em JHEP}, 02:009, 1998.
\newblock hep-th/9712251.

\bibitem{teo}
E.~Teo.
\newblock Black hole absorption cross-sections and the {A}nti-de {S}itter
  conformal field theory correspondence.
\newblock hep-th/9805014.

\bibitem{myung}
H.~Lee, N.~Kim, and Y.~Myung.
\newblock Probing the {BTZ} black hole with test fields.
\newblock hep-th/9803227.

\bibitem{cremmer79}
E.~Cremmer and B.Julia.
\newblock The {SO(8)} supergravity.
\newblock {\em Nucl. Phys. B}, 159:141--212, 1979.

\bibitem{kallosh96a}
R.~Kallosh and B.~Kol.
\newblock ${E}_7$ symmetric area of the black hole horizon.
\newblock {\em Phys.Rev.D}, 53:5344--5348, 1996.
\newblock hep-th/9602014.

\bibitem{hull96}
C.~Hull and M.~Cveti\v{c}.
\newblock Black holes and {U}-duality.
\newblock {\em Nucl.Phys.B}, 480:296--316, 1996.
\newblock hep-th/9606193.

\bibitem{freN2}
P.~Fre.
\newblock Lectures on special {K}\"{a}hler geometry and electric - magnetic
  duality rotations.
\newblock {\em Nucl.Phys.B. (Proc.Suppl)}, 45:59--114, 1996.
\newblock hep-th/9512043.

\bibitem{kallosh96b}
S.~Ferrara and R.~Kallosh.
\newblock Universality of supersymmetric attractors.
\newblock {\em Phys.Rev.D}, 54:1525--1534, 1996.
\newblock hep-th/9603090.

\bibitem{frefixed}
L.~Andrianopoli, R.~D'Auria, S.~Ferrara, P.~Fre, and M.~Trigiante.
\newblock ${E}_{7,7}$ duality, {BPS} black hole evolution and fixed scalars.
\newblock {\em Nucl.Phys.B}, 509:463--518, 1998.
\newblock hep-th/9707087.

\bibitem{cfthair1}
M.Cveti\v{c} and A.~Tseytlin.
\newblock Solitonic strings and {BPS} saturated dyonic black holes.
\newblock {\em Phys. Rev. D}, 53:5619--5633, 1996.
\newblock hep-th/9512031; Erratum-ibid. 55:3907, 1997.

\bibitem{kallosh96c}
K.~Behrndt, R.~Kallosh, J.~Rahmfeld, M.~Schmakova, and W.~Wong.
\newblock {STU} black holes and string triality.
\newblock {\em Phys.Rev.D}, 54:6293--6301, 1996.
\newblock hep-th/9608059.

\bibitem{bl98}
V.~Balasubramanian and F.~Larsen.
\newblock Near horizon geometry and black holes in four dimensions.
\newblock hep-th/9802198.

\bibitem{skenderis97a}
H.~Boonstra, B.~Peeters, and K.~Skenderis.
\newblock Duality and asymptotic geometries.
\newblock {\em Phys.Lett.B}, 411:59--67, 1997.
\newblock hep-th/9706192.

\bibitem{guna86}
M.~Gunaydin, G.~Sierra, and P.K. Townsend.
\newblock The unitary supermultiplets of {D = 3} {A}nti-de {S}itter and {D = 2}
  conformal superalgebras.
\newblock {\em Nucl.Phys.B}, 274:429, 1986.

\bibitem{gpartial}
S.~Gubser.
\newblock Can the effective string see higher partial waves?
\newblock {\em Phys.Rev.D}, 56:4984--4993, 1997.
\newblock hep-th/9704195.

\bibitem{gibbons96}
S.~Das, G.~Gibbons, and S.~Mathur.
\newblock Universality of low-energy absorption cross-sections for black holes.
\newblock {\em Phys. Rev. Lett.}, 78:417--419, 1997.
\newblock hep-th/9609052.

\bibitem{vw94}
C.~Vafa and E.~Witten.
\newblock A strong coupling test of {S}-duality.
\newblock {\em Nucl.Phys.}, 431:3--77, 1994.
\newblock hep-th/9408074.

\bibitem{toappear}
F.~Larsen and R.~Leigh.
\newblock To appear.

\end{thebibliography}

\end{document}